\documentclass[aps,prb,preprint,superscriptaddress]{revtex4-1}

\usepackage{graphicx}
\usepackage{amsmath}

\begin{document}


\title{$\pi$ - Anisotropy: A nano-carbon route to hard magnetism}


\author{Timothy Moorsom}
\email[]{T.Moorsom@leeds.ac.uk}
\affiliation{University of Leeds, Leeds, UK}

\author{Shoug Alghamdi}
\affiliation{University of Leeds, Leeds, UK}
\affiliation{Taibah University, Medina, Saudi Arabia}

\author{Sean Stansill}
\affiliation{University of Leeds, Leeds, UK}

\author{Emiliano Poli}
\affiliation{Scientific Computing Department, STFC, Rutherford Appleton Laboratory, UK.}

\author{Gilberto Teobaldi}
\affiliation{Scientific Computing Department, STFC, Rutherford Appleton Laboratory, UK.}
\affiliation{Beijing Computational Science Research Center, Beijing, China.}
\affiliation{Stephenson Institute for Renewable Energy, Department of Chemistry, University of Liverpool, UK.}
\affiliation{School of Chemistry, University of Southhampton, Southampton, UK.}

\author{Marijan Beg}
\affiliation{ Faculty of Engineering and Physical Sciences, University of Southampton, Southampton, United Kingdom.}
\affiliation{European XFEL GmbH, Holzkoppel 4, 22869 Schenefeld, Germany.}

\author{Hans Fangohr}
\affiliation{ Faculty of Engineering and Physical Sciences, University of Southampton, Southampton, United Kingdom.}
\affiliation{European XFEL GmbH, Holzkoppel 4, 22869 Schenefeld, Germany.}

\author{Matt Rogers}
\affiliation{University of Leeds, Leeds, UK}

\author{Zabeada Aslam}
\affiliation{University of Leeds, Leeds, UK}

\author{Mannan Ali}
\affiliation{University of Leeds, Leeds, UK}

\author{B J Hickey}
\affiliation{University of Leeds, Leeds, UK}

\author{Oscar Cespedes}
\email[]{O.Cespedes@leeds.ac.uk}
\affiliation{University of Leeds, Leeds, UK}


\date{\today}

\begin{abstract}

High coercivity magnets are an important resource for renewable energy, electric vehicles and memory technologies.   Most hard magnetic materials incorporate rare-earths such as neodymium and samarium, but the concerns about the environmental impact and supply stability of these materials is prompting research into alternatives. Here, we present a hybrid bilayer of cobalt and the nano-carbon molecule $C_{60}$ which exhibits significantly enhanced coercivity with minimal reduction in magnetisation. We demonstrate how this anisotropy enhancing effect cannot be described by existing models of molecule-metal magnetic interfaces. We outline a new form of magnetic anisotropy, arising from asymmetric magneto-electric coupling in the metal-molecule interface. Because this phenomenon arises from $\pi$ - d hybrid orbitals, we propose calling this effect $\pi$ - anisotropy. While the critical temperature of this effect is currently limited by the rotational degree of freedom of the chosen molecule, $C_{60}$, we describe how surface functionalisation would allow for the design of room-temperature, carbon based hard magnetic films.

\end{abstract}


\maketitle

\section{Introduction} \raggedbottom

The coupling between molecules and magnetic thin films has been intensively explored over the last fifteen years. It has been observed that anti-ferromagnetic interface states form between a variety of organic molecules and Co or Fe films, resulting in changes to their magnetic anisotropy. \cite{5,7,8,9}  Furthermore, it has been observed that $C_{60}$ has a profound effect on the band structure and magnetic behaviour of transition metals, inducing ferromagnetic states in otherwise non-magnetic materials. \cite{10, 11, 15}  The high electron affinity of $C_{60}$ can overcome the work function of metals such as Au, Cu and Co, leading to a transfer of spin polarised charge.\cite{12,13}  This interfacial coupling is accompanied by the formation of a polarized $\pi$-d hybrid interface state in the $C_{60}$ band gap. \cite{6} These surface interactions result in a modified density of states (DOS) at the metal surface and the formation of an anti-ferromagnetically (AF) coupled interface state detectable by transport and spectroscopy. \cite{14,16} While Co/$C_{60}$ surfaces in general exhibit increased coercivity and decreased magnetization, we observe that tuning the Co structure using a Ta seed layer leads to energy products, $\mu_0$MH, up to 8.6 $MJ/m^3$, an increase of 5.2x that of uncapped Co thin films, figure 1 a,b.

This increase cannot solely be explained by changes in DOS and interface hybridisation. The predicted change in interface anisotropy calculated by Bairagi et al in ultra-thin Co films was 1.5 meV, whereas the pinning observed in Co-$C_{60}$ films is 10.8 meV. \cite{9} However, we consider the effects of broken interfacial symmetry on in plane anisotropy. DFT simulations of the Co-$C_{60}$ interface show that the molecule adsorbs to the surface with adsorption energies of -6.5 eV when the closest carbon atom to the surface is at the vertex joining a hexagonal and pentagonal carbon ring (HP). In this orientation, the sum of all p-d hybrid bonds results in a strong out of plane electric dipole which is dependent on the in-plane magnetisation. This interfacial magneto-electric coupling explains the dramatic increase in coercivity observed in Co-$C_{60}$ films below the rotational transition of $C_{60}$.

\begin{figure}[h]
	\centering
	\includegraphics[width=1\textwidth]{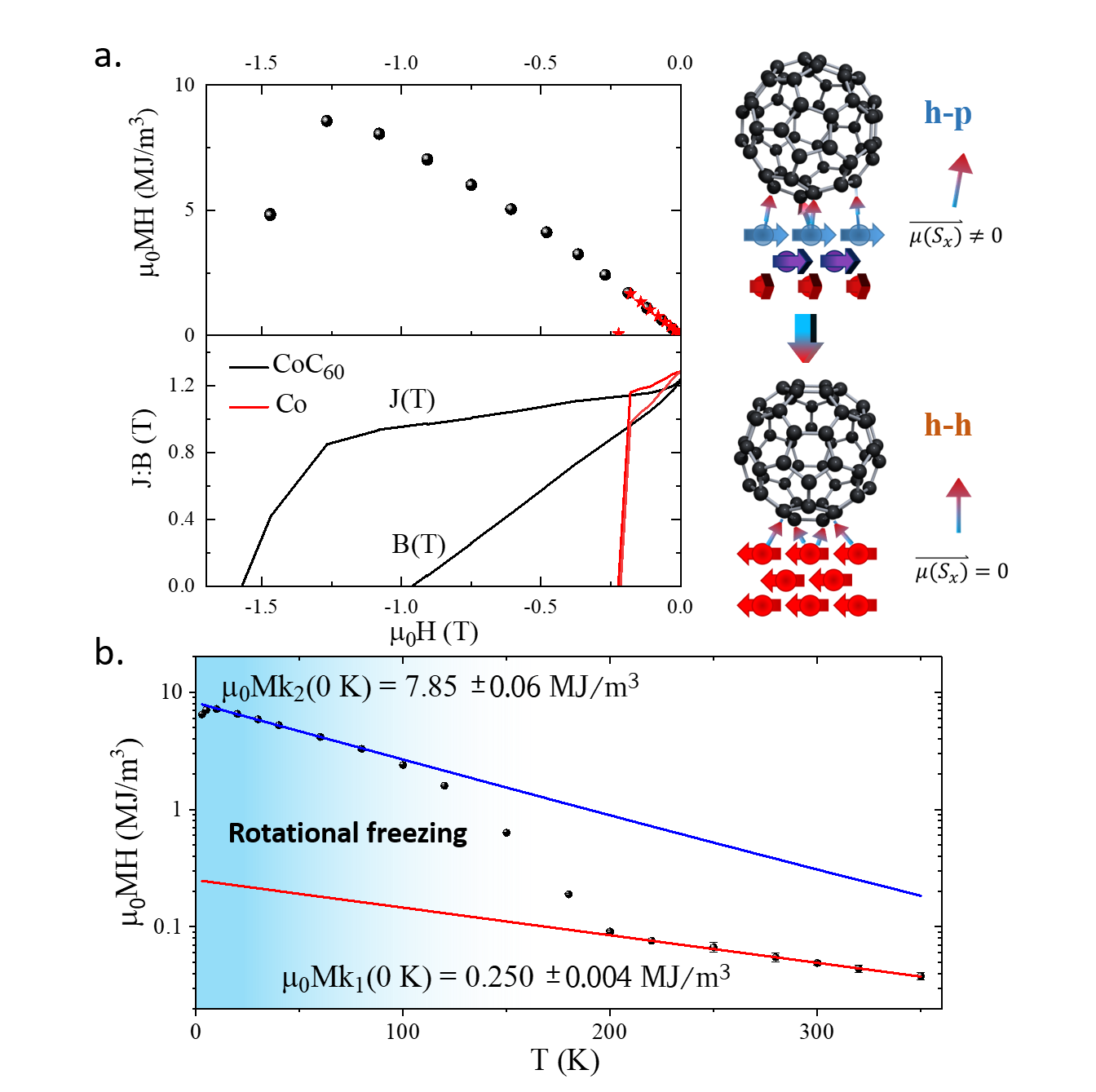} 
	\caption{a. MH and JH curves for two indentical, 3nm films of Co recorded in a SQUID-VSM after cooling in a 2 T applied field. The red points refer to a bare film of Co while the black points show a film which has been capped with 35 nm of $C_{60}$. The increase in the maximum energy product is 520 \%. The right hand images show the expected orientation of the $C_{60}$ molecule on the Co surface before demagnetisation [top] and after [bottom]. b. The energy product for a Co/$C_{60}$ film calculated as a function of temperature. There are two distinct regimes above and below the rotational transition of $C_{60}$ at 100 K. The red and blue fits are for the temperature dependent pinning factor as described in the temperature dependent Jiles-Atherton model, equation 1.\cite{18} \label{fig:figure1} }
\end{figure}

\section{Magnetometry Results}

SQUID magnetometry results show Co-$C_{60}$ bilayers cooled in an external field appear to exhibit exchange bias fields of up to 0.45 T and coercivity up to 1.5 T, figure \ref{fig:figure1}a. Exchange bias is commonly the result of coupling between ferromagnets (FM) and antiferromagnets (AF). \cite{32} While the Co/$C_{60}$ interface exhibits AF coupling, this extends only for a single monolayer and the $C_{60}$ films do not show high magnetic anisotropy or exchange coupling. Furthermore, in exchange biased FM/AF bilayers, coercivity peaks at the N\'eel temperature of the AF due to its breakdown into weakly coupled grains which contribute to domain wall pinning but not to unidirectional anisotropy. \cite{17}  However, the temperature dependent coercivity of these bilayers shows no such peak, figure \ref{fig:figure1}b. Analysis of the dependence of coercivity on temperature reveal two distinct regions, which can both be fit to a Jiles-Atherton (JA) model. \cite{18} The pinning factor is roughly equivalent to coercivity and is defined as:

\begin{equation}
k(T) = k(0)exp\left(\frac{-2T}{\beta T_c}\right)
\label{eqn:equation1}
\end{equation}

where k is the pinning factor, $\beta$ is the critical exponent of the ferromagnet and $T_c$ the Curie temperature. The high temperature region is fit to a single exponential while the low temperature region is fit to the sum of the high temperature behaviour and a low temperature pinning factor with differing k(0) and $T_c$. The high temperature pinning factor corresponds to domain wall pinning sites commonly found in thin magnetic films and described in the JA model, while the low tempature pinning factor corresponds to interfacial pinning from the $C_{60}$. The $T_c$ of the high temperature region is found to be 739 $\pm$ 6 K. While it is not possible to verify this Curie temperature in a bilayer, $C_{60}$ evaporates at between 600 – 700 K, we expect that strong hybridisation between thin-film Co and $C_{60}$ would supress $T_c$ as well as saturation magnetisation. The critical temperature of the low temperature region is found to be 351 $\pm$ 9 K, well above room temperature. The steep reduction in pinning above 100 K does not fit to a JA model but shows critical behaviour.

Following a single demagnetisation cycle, the coercivity of the loop drops by 50 \% and the exchange bias is reduced to zero. Changes in exchange bias after successive sweeps is observed in conventional exchange bias FM/AF bilayers where it is known as training. \cite {37} This is typically attributed to the movement of anti-ferromagnetic domain walls in the AF layer. However, in this case, there is no bulk AF which might allow for the formation of AF domain walls and explain the training effect. Furthermore, it is notable that only the negative coercivity changes between the first and second sweep, while the positive branch of the hysteresis loop is unchanged. This provides an alternate explanation. Rather than modelling this effect according the exchange bias model of Meiklejohn and Bean, this offset loop could be explained as the superposition of two hysteresis loops, one high coercivity and low coercivity, of which the high coercivity loop survives only a single demagnetisation cycle.

The first-order-reversal-curve (FORC) technique decomposes a hysteresis loop into individual demagnetisation quanta or hysterons. \cite{3} The distribution of hysterons in a hysteresis loop provides information about the range of activation energies for magnetisation reversal and, therefore, the variations in anisotropy, domain wall pinning and exchange bias in a thin film. This is achieved by applying a saturating positive field followed by a non-saturating reversal field, $H_f$. The sample's magnetisation is then measured while sweeping the field back to positive saturation at various field setpoints, $H_a$. This process is repeated for progressively increasing reversal fields. The hysteron density, $\rho$,  is then defined by the mixed second order differential:

\begin{equation}
\rho(H_a , H_f) = - \frac{1}{2} \frac{d^2 M}{d H_f d H_a}
\label{eqn:equation3}
\end{equation}

This can then be transformed into the bias field, $H_b$, and coercivity, $H_c$, using the definitions:

\begin{equation}
H_b = \frac{H_f - H_a}{2}
\label{eqn:equation4}
\end{equation}
\begin{equation}
H_c = \frac{H_f + H_a}{2}
\label{eqn:equation5}
\end{equation}

FORC analysis can qualitatively distinguish between a conventional exchange bias mechanism and the mechanism suggested above. In the case of conventional exchange bias with a large training effect, the change in the bias field will move the peak of the hysteron distribution closer to $H_b = 0$ while the distribution in $H_c$ will be similar or unchanged. However, if the hysteresis loop is the combination of two very different reversal mechanisms, the peak and distribution of the hysteron density will change significantly between the first and second sweeps.

\begin{figure}[h]
	\centering
	\includegraphics[width=1\textwidth]{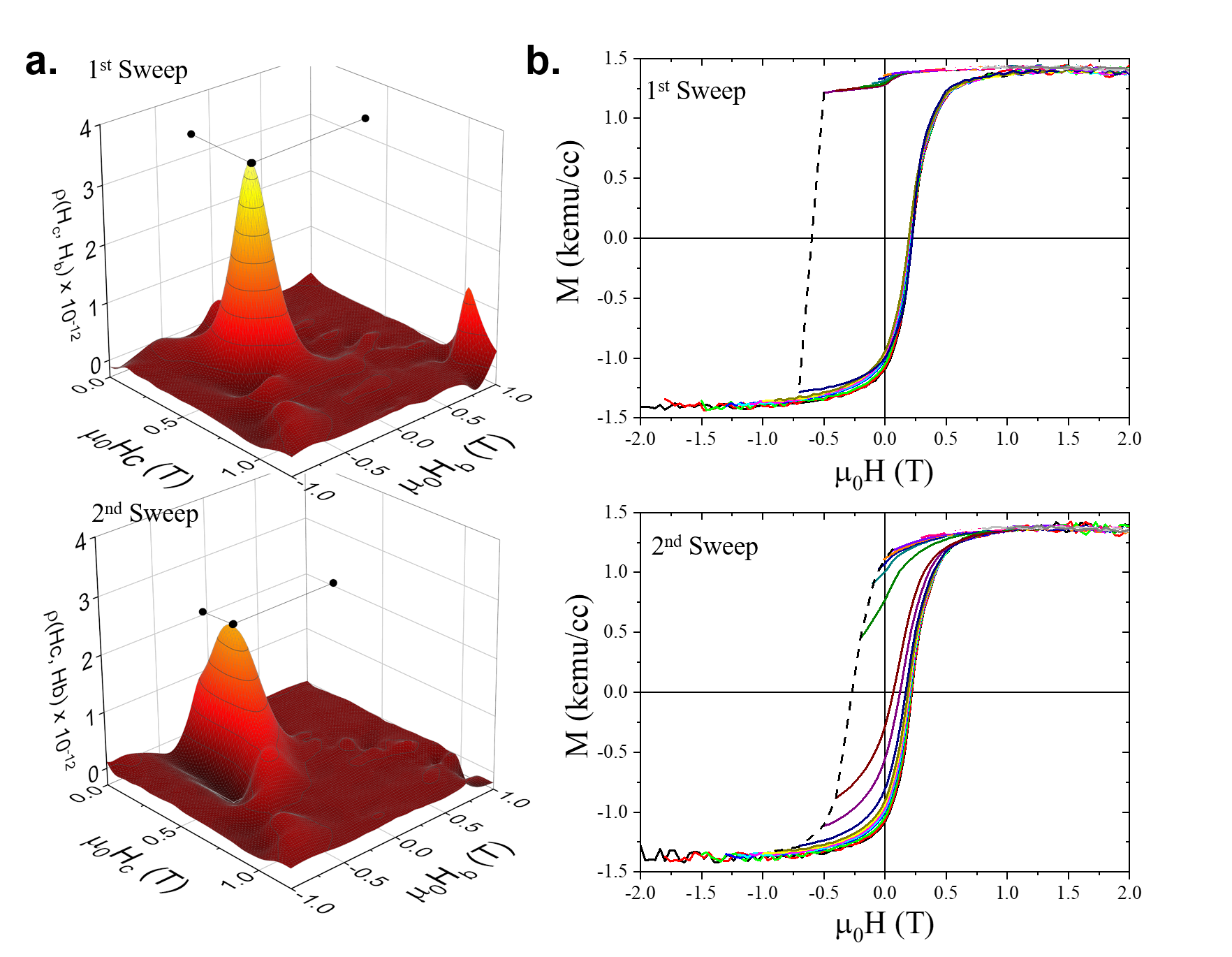} 
	\caption{a. Hysteron density plots for the first and second demagnetisation of a Co-$C_{60}$ film cooled to 5 K in a 2 T applied field. The distribution of reversal modes is significantly changed from the first to the second sweep, indicating a change in reversal mechanism rather than a training effect. This is also clear in the hysteresis loops, b. Each branch here respresents a different reversal field, $H_f$, as defined in eqn \ref{eqn:equation3}. The dotted line is a guide to the eye, showing the complete hysteresis loop. During the first sweep, demagnetisation occurs in a single, sharp step at high coercivity, while the second sweep shows a broader distribution.\label{fig:figure2} }
\end{figure}

The results of FORC measurements on a Co-$C_{60}$ bilayer during the first and second sweeps are shown in figure  \ref{fig:figure2}a. The 3D plots show the hysteron density for the first and second sweep. In the first sweep, the distribution forms a sharp peak at high bias and coercivity. Notably, the small step at zero field evident in figure \ref{fig:figure1} a does not produce a hysteron peak. This is because it is completely paramagnetic. In the second sweep, the hysteron peak is reduced, broadened and pushed towards zero bias. In addition, there is now a long tail extending to high bias and coercivity. This distribution indicates a broad range of activation energies for different reversal modes. The difference between the first and second sweep is clear in the hysteresis loops, figure \ref{fig:figure2}b.

\section{$\Pi$ - anisotropy discussion and simulation}

The transition temperature range between the high and low temperature regions in figure \ref{fig:figure1}b corresponds closely to the range over which the rotational time-scale for a $C_{60}$ molecule is changing, with the rotation being ‘frozen-out’ at 90 K in bulk films. \cite{19} STM observations of single $C_{60}$ molecules on Co, Fe and Cr surfaces reveal that the spin polarisation of the hybrid interface state is strongly dependent on interfacial symmetry, in particular in cubic metal films, where the broken interfacial symmetry gives rise to very high polarisation in the fullerenes. \cite{20, 36} Our DFT simulations show that, on the (111) plane of FCC Co, the $C_{60}$ preferentially adsorbs on the h-p vertex, or 5:6 bond, leading to -6.5 eV adsorption energy and breaking of the symmetry of the interface. This leads to a symmetry dependent interfacial spin polarisation, which varies by 0.2 $\mu_B$ between the hexagonal and pentagonal faces of the molecule (Supplemental Information, Section S4). In composites containing magnetic-transition-metals and light elements such as carbon or oxygen, spin-orbit coupling gives rise to a spin-dependence in the hybridisation between p and d orbitals. \cite{22}  Where p-d hybridisation occurs asymmetrically between multiple light atoms and a single transition metal atom, this results in a spin dependent electric dipole. 

The polarization induced by spin-dependent hybridization is defined as:

\begin{equation}
\overrightarrow{P} = \displaystyle\sum_{i,j}^{n,m} A_{ij} \left(\left|S_i\right|\left|r_{ij}\right|cos\theta_{ij}\right)^2 \hat{r}_{ij}
\label{eqn:equation2}
\end{equation}

Where $r_{ij}$ is the vector pointing from a given transition metal atom i with spin $S_i$, to a light atom j. \cite{23} The angle between the bond and the spin is given as $\theta_{ij}$. $A_{ij}$ defines the magneto-electric coupling strength. At the interface between a metal lattice comprising n bonded atoms and a molecule comprising m bonded atoms, the spin dependent contribution to the electric dipole is given by the sum of $P_{ij}$ over all bonds.

If the molecule is bonded on the vertex between two hexagonal faces, the HH orientation, all in-plane components of the polarization in equation \ref{eqn:equation2} will cancel. However, if it is bonded between a hexagonal and pentagonal face, HP orientation, there will be a component of $\Sigma S_{ij}^{xy} \cdot r_{ij}$ which does not cancel, due to the symmetry dependence of the hybridisation, meaning an in-plane spin rotation will change the magnitude of the out-of-plane electric dipole. In addition to this spin-dependent surface dipole, there exists an in-built potential between molecule and metal due to the mismatch of fermi-levels.\cite{12} The interaction between the spin dependent dipole and in-built potential adds a new spin-dependent electrostatic term to the anisotropy of the Co surface. 

\begin{figure}[h]
	\centering
	\includegraphics[width=1\textwidth]{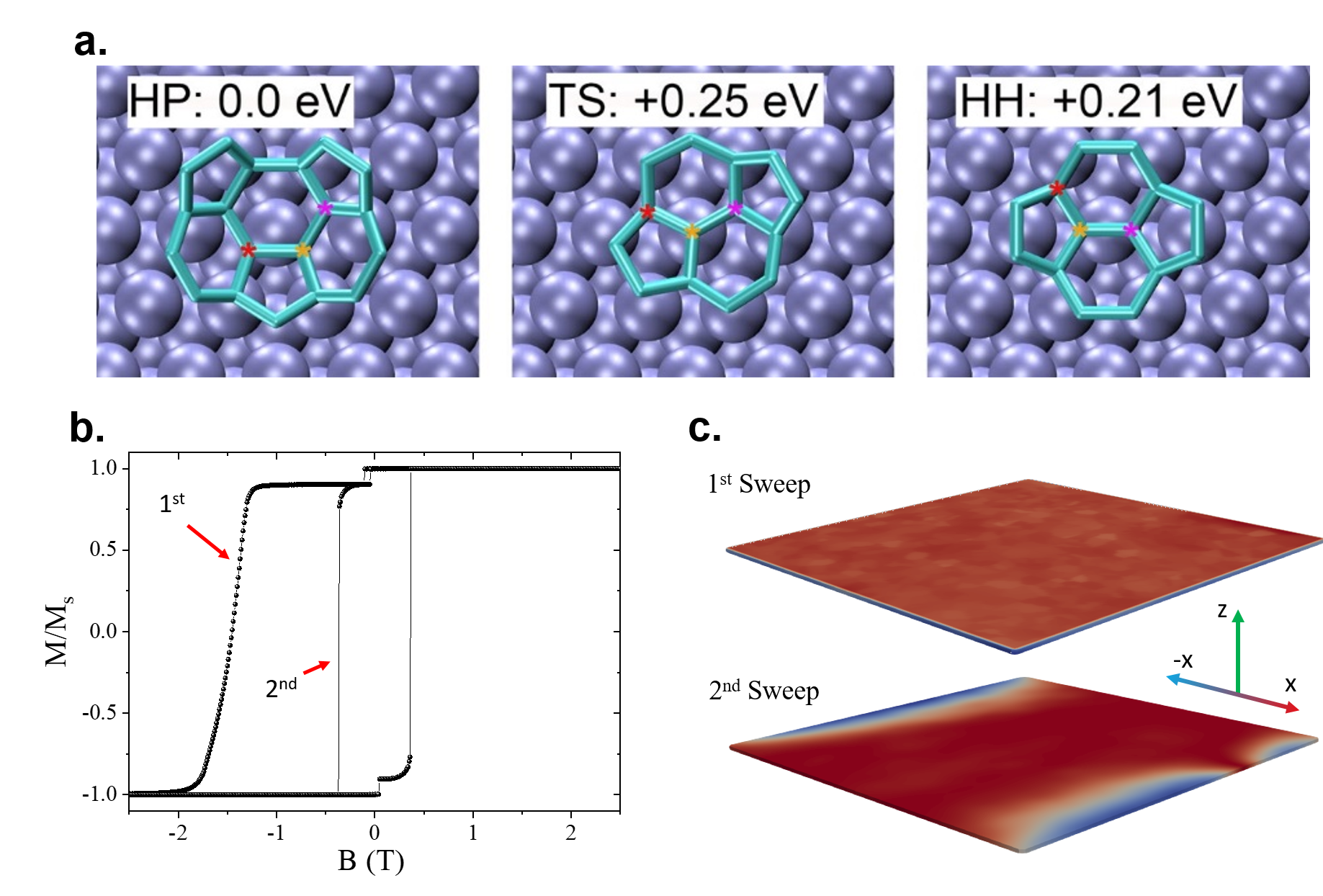} 
	\caption{ a. Representations of the Co/$C_{60}$ stationary points during rotation as simulated via DFT. The HP configuration (left) is not symmetric with respect to the Co surface, three atoms on the hexagonal ring are marked for reference. The transition state (TS) shows the point in the HH-HP rotation where the surface energy is maximized. This is a first approximation to the energy barrier which must be overcome by the magneto-electric torque at the interface. The C60 will then reach the meta-stable HH state. b. Hysteresis loop simulated using the MuMax3 code. The first sweep includes an anti-ferromagnetic surface layer with very high anisotoropy (K = 27 $MJ/m^3$) and a paramagnetic nucleation layer representing the Ta/Co intermixing region observed in cross-sectional TEM (supplemental information S1). The surface pinning replicates the vertical domain wall nucleation predicted in the $\pi$ – anisotropy model. In the second sweep, this surface anisotropy is reduced, resulting in the formation of in-plane domains. c. Shows a colour map for the slab simulated in b. Red indicates spins pointing in the +x direction and blue in the –x direction. During the first sweep, the surface remains entirely pinned while a vertical domain is compressed toward the interface. In the second sweep, the lack of strong surface pinning allows in-plane domains to form, lowering the coercivity. \label{fig:figure3} }
\end{figure}

This magneto-electric coupling means a rotation of the in-plane spins will exert a torque on the $C_{60}$ molecule. The observed surface exchange energy density at 3 K is 10.8 meV and the thermal energy corresponding to the centre of the transition in figure 1b is 12.8 meV. DFT predicts an interfacial dipole density between a 4x4 Co(111) slab and a $C_{60}$ molecule of $3.79 \times 10^{-3} e/\AA$ for the h-p configuration. The magnitude of the spin-dependent dipole is dependent on the magneto-electric coupling, $A_{ij}$, of Co/$C_{60}$, which is currently unknown. However, using example values for cobalt-ferrite gives a change in the spin dependent dipole density of approximately $1 \times 10^{-6} e/\AA$ for a 90$^{\circ}$ rotation of the surface spins of the 4x4 Co slab. \cite{30} This estimate assumes an average bond length of 0.14 nm and ignores any structural relaxation of the Co surface. This change in interfacial dipole density means that there is an electrostatic barrier of 10 - 100 meV preventing the surface magnetisation from rotating in-plane. Despite the very high adsorption energy of the $C_{60}$ molecule on the Co surface, DFT simulations of the transition state (TS) predict a maximum energy barrier to rotation from HP to HH of 0.25 eV, figure \ref{fig:figure3}a. This demonstrates that the energy required to rotate the $C_{60}$ molecule on the surface is signficicantly lower than the adsoprtion energy and is likely to be further reduced in real systems due to surface defects. We predict that the surface $C_{60}$ molecules undergo a rotation from the HP to the meta-stable HH configuration due to the magneto-electric torque exerted on the cage by the rotation of the surface magnetisation of the Co film. Such spin dependent distortions have been observed in molecule-metal interfaces using molecules such as Pentacene. \cite{41}

Once rotated, the symmetry of the HH configuration means there will be no magneto-electric torque to rotate the molecule back into the HP configuration due to the higher symmetry of this orientation. The barrier for the meta-stable HH configuration is found from DFT to be 40 meV. While this is also likely to be lower in a real surface, this explains why the exchange bias cannot be restored without heating the bilayer above its transition temperature. The superposition of high and low coercivity loops produces a similar effect to the training observed in AF/FM exchange biased bilayers, except without any actual unidirectional anisotropy. \cite{20} This model predicts an ideal surface energy density of 32 $mJ/m^2$ as compared to 0.9 $mJ/m^2$ predicted in Co/IrMn. \cite{29} This explains how a molecule-metal bilayer is able to produce a bias field 15x greater than that observed in Co/IrMn despite the weak interactions between magnetic molecules. \cite{17} This also explains the unexpected temperature dependence and magnitude of this effect both as observed in Co-$C_{60}$, and in previous studies of molecular exchange bias. \cite{21}  Because this form of anisotropy arises from spin dependent hybridization of molecular $\pi$ orbitals, we propose calling this effect $\pi$-anisotropy.

We performed micromagnetic simulations of a Co film in contact with an antiferromagnetic layer with anisotropy K = 27 $MJ/m^3$ which simulates the surface pinning. The bottom surface of the Co film is in contact with a 3 nm paramagnetic layer which simulates a Ta/Co intermixing region. The hysteresis simulation is initialised in the positive x-direction and relaxed in a 2.5 T field to simulate field cooling, varying the external magnetic field between 2.5 and -2.5 T in steps of 10 mT. We relax the system to an equilibrium state at each value of an external magnetic field and use the resulting configuration as an initial state for a new energy minimization. These simulations show coercivity of 1.5 T, figure \ref{fig:figure3}b. When the Co slab is saturated in the –x direction, the anisotropy of the surface pinning layer is reduced to K = 1 $MJ/m^3$ and exchange stiffness A = 4 $pJ/m$. This simulates the depinning of the surface due to the rotation of the molecules into the symmetric HH configuration. As a result, the sweep from -2.5 T to +2.5 T gives a coercivity of only 0.3 T and no vertical domain wall formation is observed. 3D plots of the vertical and lateral domain wall formation in the two cases are shown in figure \ref{fig:figure3}c. The full simulated loop replicates that observed experimentally despite the simulation having no unidirectional anisotropy.

\section{Transport} \raggedbottom

\begin{figure}[h]
	\centering
	\includegraphics[width=1\textwidth]{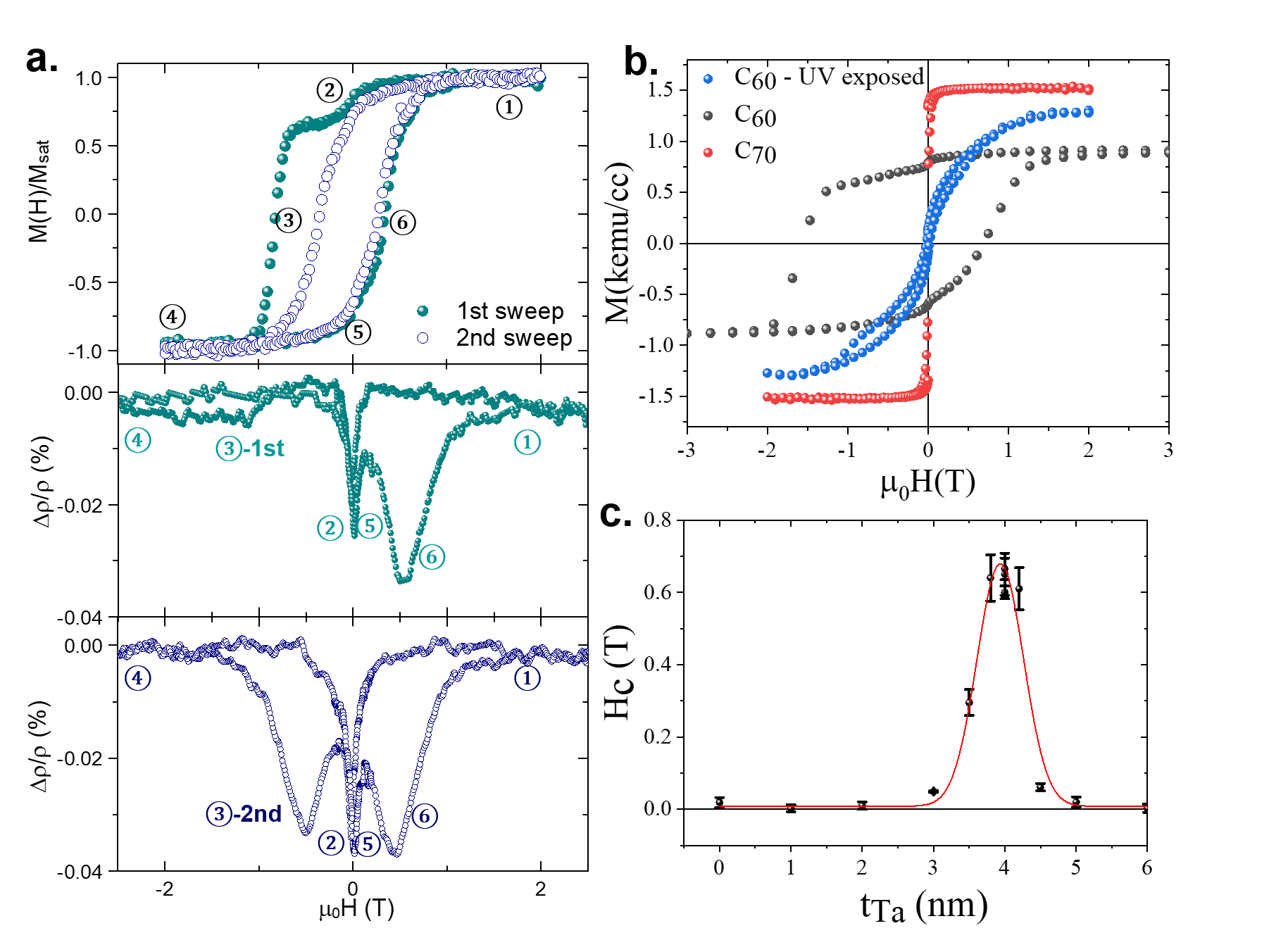} 
	\caption{a. [Top], hysteresis loops obtained from the first and second sweeps after cooling a Co/$C_{60}$ bilayer in a 2 T applied field to 5 K. The exchange bias and asymmetry is completely destroyed after a single cycle. [Middle], AMR (anisotropic magneto-resistance) recorded during the first sweep. Note that the magnetoresistance, $\Delta \rho$, is zero at point 3. [Bottom], AMR recorded during second sweep. Note that the AMR now exhibits expected behaviour for both forward and backward sweeps. b. Comparison of hysteresis loops at 5 K for Co/$C_{60}$ bilayer (black), a Co/$C_{70}$ bilayer (red) and the same Co/$C_{60}$ bilayer after removing the molecular film with a combination of acetone and UV. c. Dependence of the maximum recorded coercivity at 5 K on the thickness of the Ta seed layer, showing the importance of seeding the correct structure in the Co thin film. \label{fig:figure4} }
\end{figure}

The FORC analysis and simulations both indicate the first demagnetisation occurs via an exchange spring mechansim, in which a vertical domain wall forms in the thin film which is then compressed toward a pinned interface.\cite{24} Transport measurements support this interpretation. Hysteresis loops for a Co-$C_{60}$ sample are shown in figure \ref{fig:figure4}a, with corresponding low temperature transport data. The in-plane magnetoresistance was measured after cooling to 5K in a 2T applied field and performing two consecutive demagnetisation sweeps. The reversible step, point 2, corresponds to the formation of a vertical domain wall (DW), which is compressed toward the Co/$C_{60}$ interface with increasing field. When the molecules rotate, the vertical domain wall sweeps coherently across the film producing a sharp peak in the hysteron density, figure \ref{fig:figure4}a point 3. After the first demagnetisation, this two step reversal no longer occurs and there is a broader distribution of reversal modes, figure \ref{fig:figure4}a point 6. Anisotropic magnetoresistance (AMR) measurements show a negative peak at zero field, while the first demagnetization at higher field does not feature in the AMR at all. After de-pinning, however, negative peaks are observed in the high field AMR for both forward and backward sweeps indicating reversal through the formation of lateral domain walls. In magnetic thin films at low temperature, negative AMR is strongly dependent on spin scattering at domain walls, making the MR an approximate probe of the density of in plane domain walls. \cite{39} The lack of any change in MR during the first reversal indicates this reversal does not occur through the formation of in-plane domain walls. Molecular exchange bias has previously been observed to lead to asymmetric, negative MR in thin Co films but the explanation has until now been elusive. \cite{25}

Removing $C_{60}$ from the Co surface or using a molecule with a different symmetry does not lead to pinning. A comparison between a Co-$C_{60}$ bilayer, a Co-$C_{70}$ bilayer and a Co-$C_{60}$ layer in which the molecules have been removed using a solvent and UV exposure process is shown in figure \ref{fig:figure4}b. The cleaning process used to remove the molecules is summarised in the supplemental information, section S3. Removing $C_{60}$ from the surface results in a 98\% drop in coercivity and complete removal of the exchange bias. The use of $C_{70}$ in place of $C_{60}$, which is chemically almost identical but has lower symmetry, results in no pinning. Similarly, changing the structure of the Co thin film has a strong effect on the coercivity. The roughness, crystal structure and orientation of the Co surface is strongly dependent on the seed layer. In order to produce high pinning, the Ta seed layer must be in a 1 nm thickness window, figure \ref{fig:figure4}c. These limitations are consistent with the predictions of DFT. They also suggest that further research on the ideal surface properties may increase the critical temperature and magnitude of this effect.

\section{Conclusion} \raggedbottom

We have measured the properties of Co/$C_{60}$ bilayers and demonstrated an extremely strong anisotropy enhancing effect arising from the $C_{60}$ film. We have demonstrated how this anisotropy enhancing effect and resulting loop asymmetry cannot be explained by conventional models of exchange bias and surface anisotropy, indicating that molecular exchange bias is a distinct phenomenon. Non-magnetic $C_{60}$ is responsible for exchange spring-like behaviour through $\pi$-d hybridisation at the interface producing a spin dependent surface dipole which interacts with the in-built potential to create a new form of surface or $\pi$-anisotropy. Because this phenomenon would theoretically require only a single molecular layer to pin thin metal films, bilayers of this type may represent a means to create thin films and multi-layers with extremely high $\mu_0 MH$ energy products. As of yet, this phenomenon is limited to low temperatures. However, we have shown evidence that the critical temperature could be much higher if it were not for the rotational degree of freedom in $C_{60}$. A better choice of molecule, with reduced symmetry, dopants or ligands which prevent rotation, may produce similar or even better results at higher temperatures.

\section{Acknowledgements} \raggedbottom

This work was funded by the following grants: EPSRC EP/I004483, EP/K036408, EPSRC EP/S030263/1, EP/S031081/1, EPSRC Programme grant on Skyrmionics (EP/N032128/1), The Horizon 2020 European Research Infrastructure project OpenDreamKit (676541), Taibah University PhD Scholarship.
This work made use of the ARCHER (via the UKCP Consortium, EPSRC UK EP/P022189/1 and EP/P022189/2), UK Materials and Molecular Modelling Hub (EPSRC UK EP/P020194/1) and STFC Scientific Computing Department's SCARF High-Performance Computing facilities.

\section{Appendix A: Methods}

Magnetometry was performed in a Quantum-Design MPMS3 SQUID-VSM with samples mounted on single-crystal quartz paddles. Samples were cooled from 350 K in a 2T field applied by a superconducting solenoid. Unless otherwise stated, energy product calculations and coercive fields refer to the first sweep from +2T to -2T. Electrical transport was performed at 5 K in an Oxford Instruments cryostat. Films of Co/$C_{60}$ were bonded using Al wire. $C_{60}$ films were cleaned by soaking in acetone for 90 minutes, drying and then exposing to a Xe-Hg arc lamp. Raman and photoluminescence measurements were used to determine when the molecular film had been completely removed.

We performed finite-element micromagnetic simulations using Finmag \cite{26} and finite-difference simulations using Mumax3 (stable release 3.10) on a portion of a 3 nm thin Co film split into 128x128x10 cells. Co film is given bulk values for anisotropy, K = 60 $kJ/m^3$ exchange stiffness, A = 30 $pJ/m$, and magnetization, M = 1400 $emu/cc$. The antiferromagnetic interface anisotropy barrier is 27 $MJ/m^3$. Values were chosen to match the simulated coercivity to experimental data. DFT simulations were performed using the Projected Augmented Wave (PAW) formalism as implemented in the VASP program. \cite{38} This is detailed in the supplemental information, section S4.

\bibliography{Permag_final_PRB}

\begin{thebibliography}{31}%
\makeatletter
\providecommand \@ifxundefined [1]{%
 \@ifx{#1\undefined}
}%
\providecommand \@ifnum [1]{%
 \ifnum #1\expandafter \@firstoftwo
 \else \expandafter \@secondoftwo
 \fi
}%
\providecommand \@ifx [1]{%
 \ifx #1\expandafter \@firstoftwo
 \else \expandafter \@secondoftwo
 \fi
}%
\providecommand \natexlab [1]{#1}%
\providecommand \enquote  [1]{``#1''}%
\providecommand \bibnamefont  [1]{#1}%
\providecommand \bibfnamefont [1]{#1}%
\providecommand \citenamefont [1]{#1}%
\providecommand \href@noop [0]{\@secondoftwo}%
\providecommand \href [0]{\begingroup \@sanitize@url \@href}%
\providecommand \@href[1]{\@@startlink{#1}\@@href}%
\providecommand \@@href[1]{\endgroup#1\@@endlink}%
\providecommand \@sanitize@url [0]{\catcode `\\12\catcode `\$12\catcode
  `\&12\catcode `\#12\catcode `\^12\catcode `\_12\catcode `\%12\relax}%
\providecommand \@@startlink[1]{}%
\providecommand \@@endlink[0]{}%
\providecommand \url  [0]{\begingroup\@sanitize@url \@url }%
\providecommand \@url [1]{\endgroup\@href {#1}{\urlprefix }}%
\providecommand \urlprefix  [0]{URL }%
\providecommand \Eprint [0]{\href }%
\providecommand \doibase [0]{http://dx.doi.org/}%
\providecommand \selectlanguage [0]{\@gobble}%
\providecommand \bibinfo  [0]{\@secondoftwo}%
\providecommand \bibfield  [0]{\@secondoftwo}%
\providecommand \translation [1]{[#1]}%
\providecommand \BibitemOpen [0]{}%
\providecommand \bibitemStop [0]{}%
\providecommand \bibitemNoStop [0]{.\EOS\space}%
\providecommand \EOS [0]{\spacefactor3000\relax}%
\providecommand \BibitemShut  [1]{\csname bibitem#1\endcsname}%
\let\auto@bib@innerbib\@empty
\bibitem [{\citenamefont {Cespedes}\ \emph {et~al.}(2004)\citenamefont
  {Cespedes}, \citenamefont {Ferreira}, \citenamefont {S~Sanvito},\ and\
  \citenamefont {Coey}}]{5}%
  \BibitemOpen
  \bibfield  {author} {\bibinfo {author} {\bibfnamefont {O.}~\bibnamefont
  {Cespedes}}, \bibinfo {author} {\bibfnamefont {M.~S.}\ \bibnamefont
  {Ferreira}}, \bibinfo {author} {\bibfnamefont {M.~K.}\ \bibnamefont
  {S~Sanvito}}, \ and\ \bibinfo {author} {\bibfnamefont {J.~M.~D.}\
  \bibnamefont {Coey}},\ }\href@noop {} {\bibfield  {journal} {\bibinfo
  {journal} {Journal of Physics Condensed Matter}\ }\textbf {\bibinfo {volume}
  {16}} (\bibinfo {year} {2004})}\BibitemShut {NoStop}%
\bibitem [{\citenamefont {Sanvito}(2010)}]{7}%
  \BibitemOpen
  \bibfield  {author} {\bibinfo {author} {\bibfnamefont {S.}~\bibnamefont
  {Sanvito}},\ }\href@noop {} {\bibfield  {journal} {\bibinfo  {journal}
  {Nature Physics}\ }\textbf {\bibinfo {volume} {6}} (\bibinfo {year}
  {2010})}\BibitemShut {NoStop}%
\bibitem [{\citenamefont {Moorsom}\ \emph {et~al.}(2014)\citenamefont
  {Moorsom}, \citenamefont {Wheeler}, \citenamefont {Khan}, \citenamefont
  {Ma'Mari}, \citenamefont {Kinane}, \citenamefont {Langridge}, \citenamefont
  {d~Ciudad}, \citenamefont {Bedoya-Pinto}, \citenamefont {Hueso},
  \citenamefont {Teobaldi}, \citenamefont {Lazarov}, \citenamefont {Gilks},
  \citenamefont {Burnell}, \citenamefont {Hickey},\ and\ \citenamefont
  {Cespedes}}]{8}%
  \BibitemOpen
  \bibfield  {author} {\bibinfo {author} {\bibfnamefont {T.}~\bibnamefont
  {Moorsom}}, \bibinfo {author} {\bibfnamefont {M.}~\bibnamefont {Wheeler}},
  \bibinfo {author} {\bibfnamefont {T.~M.}\ \bibnamefont {Khan}}, \bibinfo
  {author} {\bibfnamefont {F.~A.}\ \bibnamefont {Ma'Mari}}, \bibinfo {author}
  {\bibfnamefont {C.}~\bibnamefont {Kinane}}, \bibinfo {author} {\bibfnamefont
  {S.}~\bibnamefont {Langridge}}, \bibinfo {author} {\bibnamefont {d~Ciudad}},
  \bibinfo {author} {\bibfnamefont {A.}~\bibnamefont {Bedoya-Pinto}}, \bibinfo
  {author} {\bibfnamefont {L.}~\bibnamefont {Hueso}}, \bibinfo {author}
  {\bibfnamefont {G.}~\bibnamefont {Teobaldi}}, \bibinfo {author}
  {\bibfnamefont {V.~K.}\ \bibnamefont {Lazarov}}, \bibinfo {author}
  {\bibfnamefont {D.}~\bibnamefont {Gilks}}, \bibinfo {author} {\bibfnamefont
  {G.}~\bibnamefont {Burnell}}, \bibinfo {author} {\bibfnamefont {B.~J.}\
  \bibnamefont {Hickey}}, \ and\ \bibinfo {author} {\bibfnamefont
  {O.}~\bibnamefont {Cespedes}},\ }\href@noop {} {\bibfield  {journal}
  {\bibinfo  {journal} {Physical Review B}\ }\textbf {\bibinfo {volume} {90}}
  (\bibinfo {year} {2014})}\BibitemShut {NoStop}%
\bibitem [{\citenamefont {Bairagi}\ \emph {et~al.}(2015)\citenamefont
  {Bairagi}, \citenamefont {Bellec}, \citenamefont {Repain}, \citenamefont
  {Chacon}, \citenamefont {Girard}, \citenamefont {Garreau}, \citenamefont
  {Lagoute}, \citenamefont {Rousset}, \citenamefont {Brietwieser},
  \citenamefont {Hu}, \citenamefont {Chao}, \citenamefont {Pai}, \citenamefont
  {Smogunov},\ and\ \citenamefont {Barreteau}}]{9}%
  \BibitemOpen
  \bibfield  {author} {\bibinfo {author} {\bibfnamefont {K.}~\bibnamefont
  {Bairagi}}, \bibinfo {author} {\bibfnamefont {A.}~\bibnamefont {Bellec}},
  \bibinfo {author} {\bibfnamefont {V.}~\bibnamefont {Repain}}, \bibinfo
  {author} {\bibfnamefont {C.}~\bibnamefont {Chacon}}, \bibinfo {author}
  {\bibfnamefont {Y.}~\bibnamefont {Girard}}, \bibinfo {author} {\bibfnamefont
  {Y.}~\bibnamefont {Garreau}}, \bibinfo {author} {\bibfnamefont
  {J.}~\bibnamefont {Lagoute}}, \bibinfo {author} {\bibfnamefont
  {S.}~\bibnamefont {Rousset}}, \bibinfo {author} {\bibfnamefont
  {P.}~\bibnamefont {Brietwieser}}, \bibinfo {author} {\bibfnamefont {Y.-C.}\
  \bibnamefont {Hu}}, \bibinfo {author} {\bibfnamefont {Y.~C.}\ \bibnamefont
  {Chao}}, \bibinfo {author} {\bibfnamefont {W.~W.}\ \bibnamefont {Pai}},
  \bibinfo {author} {\bibfnamefont {A.}~\bibnamefont {Smogunov}}, \ and\
  \bibinfo {author} {\bibfnamefont {C.}~\bibnamefont {Barreteau}},\ }\href@noop
  {} {\bibfield  {journal} {\bibinfo  {journal} {Physical Review Letters}\
  }\textbf {\bibinfo {volume} {114}} (\bibinfo {year} {2015})}\BibitemShut
  {NoStop}%
\bibitem [{\citenamefont {Ma'Mari}\ \emph {et~al.}(2015)\citenamefont
  {Ma'Mari}, \citenamefont {Moorsom}, \citenamefont {Teobaldi}, \citenamefont
  {Deacon}, \citenamefont {Prokscha}, \citenamefont {Luetkens}, \citenamefont
  {Lee}, \citenamefont {Sterbinsky}, \citenamefont {Arena}, \citenamefont
  {MacLaren}, \citenamefont {Flokstra}, \citenamefont {Ali}, \citenamefont
  {Wheeler}, \citenamefont {Burnell}, \citenamefont {Hickey},\ and\
  \citenamefont {Cespedes}}]{10}%
  \BibitemOpen
  \bibfield  {author} {\bibinfo {author} {\bibfnamefont {F.~A.}\ \bibnamefont
  {Ma'Mari}}, \bibinfo {author} {\bibfnamefont {T.}~\bibnamefont {Moorsom}},
  \bibinfo {author} {\bibfnamefont {G.}~\bibnamefont {Teobaldi}}, \bibinfo
  {author} {\bibfnamefont {W.}~\bibnamefont {Deacon}}, \bibinfo {author}
  {\bibfnamefont {T.}~\bibnamefont {Prokscha}}, \bibinfo {author}
  {\bibfnamefont {H.}~\bibnamefont {Luetkens}}, \bibinfo {author}
  {\bibfnamefont {S.}~\bibnamefont {Lee}}, \bibinfo {author} {\bibfnamefont
  {G.~E.}\ \bibnamefont {Sterbinsky}}, \bibinfo {author} {\bibfnamefont
  {D.~A.}\ \bibnamefont {Arena}}, \bibinfo {author} {\bibfnamefont {D.~A.}\
  \bibnamefont {MacLaren}}, \bibinfo {author} {\bibfnamefont {M.}~\bibnamefont
  {Flokstra}}, \bibinfo {author} {\bibfnamefont {M.}~\bibnamefont {Ali}},
  \bibinfo {author} {\bibfnamefont {M.~C.}\ \bibnamefont {Wheeler}}, \bibinfo
  {author} {\bibfnamefont {G.}~\bibnamefont {Burnell}}, \bibinfo {author}
  {\bibfnamefont {B.~J.}\ \bibnamefont {Hickey}}, \ and\ \bibinfo {author}
  {\bibfnamefont {O.}~\bibnamefont {Cespedes}},\ }\href@noop {} {\bibfield
  {journal} {\bibinfo  {journal} {Nature}\ }\textbf {\bibinfo {volume} {524}}
  (\bibinfo {year} {2015})}\BibitemShut {NoStop}%
\bibitem [{\citenamefont {Al~Ma'Mari}\ \emph {et~al.}(2017)\citenamefont
  {Al~Ma'Mari}, \citenamefont {Rogers}, \citenamefont {Alghamdi}, \citenamefont
  {Moorsom}, \citenamefont {Lee}, \citenamefont {Prokscha}, \citenamefont
  {Luetkens}, \citenamefont {Valvidares}, \citenamefont {Teobaldi},
  \citenamefont {Flokstra}, \citenamefont {Stewart}, \citenamefont {Gargiani},
  \citenamefont {Ali}, \citenamefont {Burnell}, \citenamefont {Hickey},\ and\
  \citenamefont {Cespedes}}]{11}%
  \BibitemOpen
  \bibfield  {author} {\bibinfo {author} {\bibfnamefont {F.}~\bibnamefont
  {Al~Ma'Mari}}, \bibinfo {author} {\bibfnamefont {M.}~\bibnamefont {Rogers}},
  \bibinfo {author} {\bibfnamefont {S.}~\bibnamefont {Alghamdi}}, \bibinfo
  {author} {\bibfnamefont {T.}~\bibnamefont {Moorsom}}, \bibinfo {author}
  {\bibfnamefont {S.}~\bibnamefont {Lee}}, \bibinfo {author} {\bibfnamefont
  {T.}~\bibnamefont {Prokscha}}, \bibinfo {author} {\bibfnamefont
  {H.}~\bibnamefont {Luetkens}}, \bibinfo {author} {\bibfnamefont
  {M.}~\bibnamefont {Valvidares}}, \bibinfo {author} {\bibfnamefont
  {G.}~\bibnamefont {Teobaldi}}, \bibinfo {author} {\bibfnamefont
  {M.}~\bibnamefont {Flokstra}}, \bibinfo {author} {\bibfnamefont
  {R.}~\bibnamefont {Stewart}}, \bibinfo {author} {\bibfnamefont
  {P.}~\bibnamefont {Gargiani}}, \bibinfo {author} {\bibfnamefont
  {M.}~\bibnamefont {Ali}}, \bibinfo {author} {\bibfnamefont {G.}~\bibnamefont
  {Burnell}}, \bibinfo {author} {\bibfnamefont {B.~J.}\ \bibnamefont {Hickey}},
  \ and\ \bibinfo {author} {\bibfnamefont {O.}~\bibnamefont {Cespedes}},\
  }\href {\doibase 10.1073/pnas.1620216114} {\bibfield  {journal} {\bibinfo
  {journal} {Proceedings of the National Academy of Sciences}\ }\textbf
  {\bibinfo {volume} {114}},\ \bibinfo {pages} {5583} (\bibinfo {year}
  {2017})},\ \Eprint
  {http://arxiv.org/abs/https://www.pnas.org/content/114/22/5583.full.pdf}
  {https://www.pnas.org/content/114/22/5583.full.pdf} \BibitemShut {NoStop}%
\bibitem [{\citenamefont {Martín-Olivera}\ \emph {et~al.}(2017)\citenamefont
  {Martín-Olivera}, \citenamefont {Shchukin},\ and\ \citenamefont
  {Teobaldi}}]{15}%
  \BibitemOpen
  \bibfield  {author} {\bibinfo {author} {\bibfnamefont {L.}~\bibnamefont
  {Martín-Olivera}}, \bibinfo {author} {\bibfnamefont {D.~G.}\ \bibnamefont
  {Shchukin}}, \ and\ \bibinfo {author} {\bibfnamefont {G.}~\bibnamefont
  {Teobaldi}},\ }\href {\doibase 10.1021/acs.jpcc.7b08476} {\bibfield
  {journal} {\bibinfo  {journal} {The Journal of Physical Chemistry C}\
  }\textbf {\bibinfo {volume} {121}},\ \bibinfo {pages} {23777} (\bibinfo
  {year} {2017})},\ \Eprint
  {http://arxiv.org/abs/https://doi.org/10.1021/acs.jpcc.7b08476}
  {https://doi.org/10.1021/acs.jpcc.7b08476} \BibitemShut {NoStop}%
\bibitem [{\citenamefont {Lu}\ \emph {et~al.}(2004)\citenamefont {Lu},
  \citenamefont {Grobis}, \citenamefont {Khoo}, \citenamefont {Louie},\ and\
  \citenamefont {Crommie}}]{12}%
  \BibitemOpen
  \bibfield  {author} {\bibinfo {author} {\bibfnamefont {X.}~\bibnamefont
  {Lu}}, \bibinfo {author} {\bibfnamefont {M.}~\bibnamefont {Grobis}}, \bibinfo
  {author} {\bibfnamefont {K.~H.}\ \bibnamefont {Khoo}}, \bibinfo {author}
  {\bibfnamefont {S.~G.}\ \bibnamefont {Louie}}, \ and\ \bibinfo {author}
  {\bibfnamefont {M.~F.}\ \bibnamefont {Crommie}},\ }\href {\doibase
  10.1103/PhysRevB.70.115418} {\bibfield  {journal} {\bibinfo  {journal} {Phys.
  Rev. B}\ }\textbf {\bibinfo {volume} {70}},\ \bibinfo {pages} {115418}
  (\bibinfo {year} {2004})}\BibitemShut {NoStop}%
\bibitem [{\citenamefont {Sau}\ \emph {et~al.}(2008)\citenamefont {Sau},
  \citenamefont {Neaton}, \citenamefont {Choi}, \citenamefont {Louie},\ and\
  \citenamefont {Cohen}}]{13}%
  \BibitemOpen
  \bibfield  {author} {\bibinfo {author} {\bibfnamefont {J.~D.}\ \bibnamefont
  {Sau}}, \bibinfo {author} {\bibfnamefont {J.~B.}\ \bibnamefont {Neaton}},
  \bibinfo {author} {\bibfnamefont {H.~J.}\ \bibnamefont {Choi}}, \bibinfo
  {author} {\bibfnamefont {S.~G.}\ \bibnamefont {Louie}}, \ and\ \bibinfo
  {author} {\bibfnamefont {M.~L.}\ \bibnamefont {Cohen}},\ }\href {\doibase
  10.1103/PhysRevLett.101.026804} {\bibfield  {journal} {\bibinfo  {journal}
  {Phys. Rev. Lett.}\ }\textbf {\bibinfo {volume} {101}},\ \bibinfo {pages}
  {026804} (\bibinfo {year} {2008})}\BibitemShut {NoStop}%
\bibitem [{\citenamefont {Barraud}\ \emph {et~al.}(2010)\citenamefont
  {Barraud}, \citenamefont {Seneor}, \citenamefont {Mattana}, \citenamefont
  {Fusil}, \citenamefont {Bouzehouane}, \citenamefont {Deranlot}, \citenamefont
  {Graziosi}, \citenamefont {Hueso}, \citenamefont {Bergenti}, \citenamefont
  {dediu}, \citenamefont {Petroff},\ and\ \citenamefont {Fert}}]{6}%
  \BibitemOpen
  \bibfield  {author} {\bibinfo {author} {\bibfnamefont {C.}~\bibnamefont
  {Barraud}}, \bibinfo {author} {\bibfnamefont {P.}~\bibnamefont {Seneor}},
  \bibinfo {author} {\bibfnamefont {R.}~\bibnamefont {Mattana}}, \bibinfo
  {author} {\bibfnamefont {S.}~\bibnamefont {Fusil}}, \bibinfo {author}
  {\bibfnamefont {K.}~\bibnamefont {Bouzehouane}}, \bibinfo {author}
  {\bibfnamefont {C.}~\bibnamefont {Deranlot}}, \bibinfo {author}
  {\bibfnamefont {P.}~\bibnamefont {Graziosi}}, \bibinfo {author}
  {\bibfnamefont {L.}~\bibnamefont {Hueso}}, \bibinfo {author} {\bibfnamefont
  {I.}~\bibnamefont {Bergenti}}, \bibinfo {author} {\bibfnamefont
  {V.}~\bibnamefont {dediu}}, \bibinfo {author} {\bibfnamefont
  {F.}~\bibnamefont {Petroff}}, \ and\ \bibinfo {author} {\bibfnamefont
  {A.}~\bibnamefont {Fert}},\ }\href@noop {} {\bibfield  {journal} {\bibinfo
  {journal} {Nature Physics}\ }\textbf {\bibinfo {volume} {6}} (\bibinfo {year}
  {2010})}\BibitemShut {NoStop}%
\bibitem [{\citenamefont {Raman}\ \emph {et~al.}(2013)\citenamefont {Raman},
  \citenamefont {Kamerbeek}, \citenamefont {Mukherjee}, \citenamefont
  {Atodiresei}, \citenamefont {Sen}, \citenamefont {Lazić}, \citenamefont
  {Caciuc}, \citenamefont {Michel}, \citenamefont {Stalke}, \citenamefont
  {Mandal}, \citenamefont {Blügel}, \citenamefont {Münzenberg},\ and\
  \citenamefont {Moodera}}]{14}%
  \BibitemOpen
  \bibfield  {author} {\bibinfo {author} {\bibfnamefont {K.~V.}\ \bibnamefont
  {Raman}}, \bibinfo {author} {\bibfnamefont {A.~M.}\ \bibnamefont
  {Kamerbeek}}, \bibinfo {author} {\bibfnamefont {A.}~\bibnamefont
  {Mukherjee}}, \bibinfo {author} {\bibfnamefont {N.}~\bibnamefont
  {Atodiresei}}, \bibinfo {author} {\bibfnamefont {T.~K.}\ \bibnamefont {Sen}},
  \bibinfo {author} {\bibfnamefont {P.}~\bibnamefont {Lazić}}, \bibinfo
  {author} {\bibfnamefont {V.}~\bibnamefont {Caciuc}}, \bibinfo {author}
  {\bibfnamefont {R.}~\bibnamefont {Michel}}, \bibinfo {author} {\bibfnamefont
  {D.}~\bibnamefont {Stalke}}, \bibinfo {author} {\bibfnamefont {S.~K.}\
  \bibnamefont {Mandal}}, \bibinfo {author} {\bibfnamefont {S.}~\bibnamefont
  {Blügel}}, \bibinfo {author} {\bibfnamefont {M.}~\bibnamefont
  {Münzenberg}}, \ and\ \bibinfo {author} {\bibfnamefont {J.~S.}\ \bibnamefont
  {Moodera}},\ }\href@noop {} {\bibfield  {journal} {\bibinfo  {journal}
  {Nature}\ }\textbf {\bibinfo {volume} {493}},\ \bibinfo {pages} {509 }
  (\bibinfo {year} {2013})}\BibitemShut {NoStop}%
\bibitem [{\citenamefont {Djeghloul}\ \emph {et~al.}(2016)\citenamefont
  {Djeghloul}, \citenamefont {Gruber}, \citenamefont {Urbain}, \citenamefont
  {Xenioti}, \citenamefont {Joly}, \citenamefont {Boukari}, \citenamefont
  {Arabski}, \citenamefont {Bulou}, \citenamefont {Scheurer}, \citenamefont
  {Bertran}, \citenamefont {Le~Fèvre}, \citenamefont {Taleb-Ibrahimi},
  \citenamefont {Wulfhekel}, \citenamefont {Garreau}, \citenamefont
  {Hajjar-Garreau}, \citenamefont {Wetzel}, \citenamefont {Alouani},
  \citenamefont {Beaurepaire}, \citenamefont {Bowen},\ and\ \citenamefont
  {Weber}}]{16}%
  \BibitemOpen
  \bibfield  {author} {\bibinfo {author} {\bibfnamefont {F.}~\bibnamefont
  {Djeghloul}}, \bibinfo {author} {\bibfnamefont {M.}~\bibnamefont {Gruber}},
  \bibinfo {author} {\bibfnamefont {E.}~\bibnamefont {Urbain}}, \bibinfo
  {author} {\bibfnamefont {D.}~\bibnamefont {Xenioti}}, \bibinfo {author}
  {\bibfnamefont {L.}~\bibnamefont {Joly}}, \bibinfo {author} {\bibfnamefont
  {S.}~\bibnamefont {Boukari}}, \bibinfo {author} {\bibfnamefont
  {J.}~\bibnamefont {Arabski}}, \bibinfo {author} {\bibfnamefont
  {H.}~\bibnamefont {Bulou}}, \bibinfo {author} {\bibfnamefont
  {F.}~\bibnamefont {Scheurer}}, \bibinfo {author} {\bibfnamefont
  {F.}~\bibnamefont {Bertran}}, \bibinfo {author} {\bibfnamefont
  {P.}~\bibnamefont {Le~Fèvre}}, \bibinfo {author} {\bibfnamefont
  {A.}~\bibnamefont {Taleb-Ibrahimi}}, \bibinfo {author} {\bibfnamefont
  {W.}~\bibnamefont {Wulfhekel}}, \bibinfo {author} {\bibfnamefont
  {G.}~\bibnamefont {Garreau}}, \bibinfo {author} {\bibfnamefont
  {S.}~\bibnamefont {Hajjar-Garreau}}, \bibinfo {author} {\bibfnamefont
  {P.}~\bibnamefont {Wetzel}}, \bibinfo {author} {\bibfnamefont
  {M.}~\bibnamefont {Alouani}}, \bibinfo {author} {\bibfnamefont
  {E.}~\bibnamefont {Beaurepaire}}, \bibinfo {author} {\bibfnamefont
  {M.}~\bibnamefont {Bowen}}, \ and\ \bibinfo {author} {\bibfnamefont
  {W.}~\bibnamefont {Weber}},\ }\href {\doibase 10.1021/acs.jpclett.6b01112}
  {\bibfield  {journal} {\bibinfo  {journal} {The Journal of Physical Chemistry
  Letters}\ }\textbf {\bibinfo {volume} {7}},\ \bibinfo {pages} {2310}
  (\bibinfo {year} {2016})},\ \bibinfo {note} {pMID: 27266579},\ \Eprint
  {http://arxiv.org/abs/https://doi.org/10.1021/acs.jpclett.6b01112}
  {https://doi.org/10.1021/acs.jpclett.6b01112} \BibitemShut {NoStop}%
\bibitem [{\citenamefont {{Raghunathan}}\ \emph {et~al.}(2009)\citenamefont
  {{Raghunathan}}, \citenamefont {{Melikhov}}, \citenamefont {{Snyder}},\ and\
  \citenamefont {{Jiles}}}]{18}%
  \BibitemOpen
  \bibfield  {author} {\bibinfo {author} {\bibfnamefont {A.}~\bibnamefont
  {{Raghunathan}}}, \bibinfo {author} {\bibfnamefont {Y.}~\bibnamefont
  {{Melikhov}}}, \bibinfo {author} {\bibfnamefont {J.~E.}\ \bibnamefont
  {{Snyder}}}, \ and\ \bibinfo {author} {\bibfnamefont {D.~C.}\ \bibnamefont
  {{Jiles}}},\ }\href {\doibase 10.1109/TMAG.2009.2022744} {\bibfield
  {journal} {\bibinfo  {journal} {IEEE Transactions on Magnetics}\ }\textbf
  {\bibinfo {volume} {45}},\ \bibinfo {pages} {3954} (\bibinfo {year}
  {2009})}\BibitemShut {NoStop}%
\bibitem [{\citenamefont {Meiklejohn}\ and\ \citenamefont {Bean}(1957)}]{32}%
  \BibitemOpen
  \bibfield  {author} {\bibinfo {author} {\bibfnamefont {W.~H.}\ \bibnamefont
  {Meiklejohn}}\ and\ \bibinfo {author} {\bibfnamefont {C.~P.}\ \bibnamefont
  {Bean}},\ }\href {\doibase 10.1103/PhysRev.105.904} {\bibfield  {journal}
  {\bibinfo  {journal} {Phys. Rev.}\ }\textbf {\bibinfo {volume} {105}},\
  \bibinfo {pages} {904} (\bibinfo {year} {1957})}\BibitemShut {NoStop}%
\bibitem [{\citenamefont {Ali}\ \emph {et~al.}(2003)\citenamefont {Ali},
  \citenamefont {Marrows},\ and\ \citenamefont {Hickey}}]{17}%
  \BibitemOpen
  \bibfield  {author} {\bibinfo {author} {\bibfnamefont {M.}~\bibnamefont
  {Ali}}, \bibinfo {author} {\bibfnamefont {C.~H.}\ \bibnamefont {Marrows}}, \
  and\ \bibinfo {author} {\bibfnamefont {B.~J.}\ \bibnamefont {Hickey}},\
  }\href {\doibase 10.1103/PhysRevB.67.172405} {\bibfield  {journal} {\bibinfo
  {journal} {Phys. Rev. B}\ }\textbf {\bibinfo {volume} {67}},\ \bibinfo
  {pages} {172405} (\bibinfo {year} {2003})}\BibitemShut {NoStop}%
\bibitem [{\citenamefont {Binek}(2004)}]{37}%
  \BibitemOpen
  \bibfield  {author} {\bibinfo {author} {\bibfnamefont {C.}~\bibnamefont
  {Binek}},\ }\href@noop {} {\bibfield  {journal} {\bibinfo  {journal}
  {Physical review B}\ }\textbf {\bibinfo {volume} {70}},\ \bibinfo {pages}
  {014421} (\bibinfo {year} {2004})}\BibitemShut {NoStop}%
\bibitem [{\citenamefont {Davies}\ \emph {et~al.}(2005)\citenamefont {Davies},
  \citenamefont {Hellwig}, \citenamefont {Fullerton}, \citenamefont {Jiang},
  \citenamefont {Bader}, \citenamefont {Zimanyi},\ and\ \citenamefont
  {Liu}}]{3}%
  \BibitemOpen
  \bibfield  {author} {\bibinfo {author} {\bibfnamefont {J.~E.}\ \bibnamefont
  {Davies}}, \bibinfo {author} {\bibfnamefont {O.}~\bibnamefont {Hellwig}},
  \bibinfo {author} {\bibfnamefont {E.~E.}\ \bibnamefont {Fullerton}}, \bibinfo
  {author} {\bibfnamefont {J.~S.}\ \bibnamefont {Jiang}}, \bibinfo {author}
  {\bibfnamefont {S.~D.}\ \bibnamefont {Bader}}, \bibinfo {author}
  {\bibfnamefont {G.~T.}\ \bibnamefont {Zimanyi}}, \ and\ \bibinfo {author}
  {\bibfnamefont {K.}~\bibnamefont {Liu}},\ }\href@noop {} {\bibfield
  {journal} {\bibinfo  {journal} {Applied Physics Letter}\ }\textbf {\bibinfo
  {volume} {86}} (\bibinfo {year} {2005})}\BibitemShut {NoStop}%
\bibitem [{\citenamefont {David}\ \emph {et~al.}(1992)\citenamefont {David},
  \citenamefont {Ibberson}, \citenamefont {Dennis}, \citenamefont {Hare},\ and\
  \citenamefont {Prassides}}]{19}%
  \BibitemOpen
  \bibfield  {author} {\bibinfo {author} {\bibfnamefont {W.~I.~F.}\
  \bibnamefont {David}}, \bibinfo {author} {\bibfnamefont {R.~M.}\ \bibnamefont
  {Ibberson}}, \bibinfo {author} {\bibfnamefont {T.~J.~S.}\ \bibnamefont
  {Dennis}}, \bibinfo {author} {\bibfnamefont {J.~P.}\ \bibnamefont {Hare}}, \
  and\ \bibinfo {author} {\bibfnamefont {K.}~\bibnamefont {Prassides}},\ }\href
  {\doibase 10.1209/0295-5075/18/8/012} {\bibfield  {journal} {\bibinfo
  {journal} {Europhysics Letters ({EPL})}\ }\textbf {\bibinfo {volume} {18}},\
  \bibinfo {pages} {735} (\bibinfo {year} {1992})}\BibitemShut {NoStop}%
\bibitem [{\citenamefont {Brems}\ \emph {et~al.}(2007)\citenamefont {Brems},
  \citenamefont {Temst},\ and\ \citenamefont {Van~Haesendonck}}]{20}%
  \BibitemOpen
  \bibfield  {author} {\bibinfo {author} {\bibfnamefont {S.}~\bibnamefont
  {Brems}}, \bibinfo {author} {\bibfnamefont {K.}~\bibnamefont {Temst}}, \ and\
  \bibinfo {author} {\bibfnamefont {C.}~\bibnamefont {Van~Haesendonck}},\
  }\href {\doibase 10.1103/PhysRevLett.99.067201} {\bibfield  {journal}
  {\bibinfo  {journal} {Phys. Rev. Lett.}\ }\textbf {\bibinfo {volume} {99}},\
  \bibinfo {pages} {067201} (\bibinfo {year} {2007})}\BibitemShut {NoStop}%
\bibitem [{\citenamefont {Li}\ \emph {et~al.}(2016)\citenamefont {Li},
  \citenamefont {Barreteau}, \citenamefont {Kawahara}, \citenamefont {Lagoute},
  \citenamefont {Chacon},\ and\ \citenamefont {Gi}}]{36}%
  \BibitemOpen
  \bibfield  {author} {\bibinfo {author} {\bibfnamefont {D.}~\bibnamefont
  {Li}}, \bibinfo {author} {\bibfnamefont {C.}~\bibnamefont {Barreteau}},
  \bibinfo {author} {\bibfnamefont {S.~L.}\ \bibnamefont {Kawahara}}, \bibinfo
  {author} {\bibfnamefont {J.}~\bibnamefont {Lagoute}}, \bibinfo {author}
  {\bibfnamefont {C.}~\bibnamefont {Chacon}}, \ and\ \bibinfo {author}
  {\bibfnamefont {Y.}~\bibnamefont {Gi}},\ }\href@noop {} {\bibfield  {journal}
  {\bibinfo  {journal} {Physical review B}\ }\textbf {\bibinfo {volume} {93}},\
  \bibinfo {pages} {085425} (\bibinfo {year} {2016})}\BibitemShut {NoStop}%
\bibitem [{\citenamefont {Murakawa}\ \emph {et~al.}(2010)\citenamefont
  {Murakawa}, \citenamefont {Onose}, \citenamefont {Miyahara}, \citenamefont
  {Furukawa},\ and\ \citenamefont {Tokura}}]{22}%
  \BibitemOpen
  \bibfield  {author} {\bibinfo {author} {\bibfnamefont {H.}~\bibnamefont
  {Murakawa}}, \bibinfo {author} {\bibfnamefont {Y.}~\bibnamefont {Onose}},
  \bibinfo {author} {\bibfnamefont {S.}~\bibnamefont {Miyahara}}, \bibinfo
  {author} {\bibfnamefont {N.}~\bibnamefont {Furukawa}}, \ and\ \bibinfo
  {author} {\bibfnamefont {Y.}~\bibnamefont {Tokura}},\ }\href {\doibase
  10.1103/PhysRevLett.105.137202} {\bibfield  {journal} {\bibinfo  {journal}
  {Phys. Rev. Lett.}\ }\textbf {\bibinfo {volume} {105}},\ \bibinfo {pages}
  {137202} (\bibinfo {year} {2010})}\BibitemShut {NoStop}%
\bibitem [{\citenamefont {Lim}\ \emph {et~al.}(2018)\citenamefont {Lim},
  \citenamefont {Saldana-Greco},\ and\ \citenamefont {Rappe}}]{23}%
  \BibitemOpen
  \bibfield  {author} {\bibinfo {author} {\bibfnamefont {J.~S.}\ \bibnamefont
  {Lim}}, \bibinfo {author} {\bibfnamefont {D.}~\bibnamefont {Saldana-Greco}},
  \ and\ \bibinfo {author} {\bibfnamefont {A.~M.}\ \bibnamefont {Rappe}},\
  }\href {\doibase 10.1103/PhysRevB.97.045115} {\bibfield  {journal} {\bibinfo
  {journal} {Phys. Rev. B}\ }\textbf {\bibinfo {volume} {97}},\ \bibinfo
  {pages} {045115} (\bibinfo {year} {2018})}\BibitemShut {NoStop}%
\bibitem [{\citenamefont {Etier}\ \emph {et~al.}(2015)\citenamefont {Etier},
  \citenamefont {Schmitz-Antoniak}, \citenamefont {Salamon}, \citenamefont
  {Trivedi}, \citenamefont {Gao}, \citenamefont {Nazrabi}, \citenamefont
  {Landers}, \citenamefont {Gautam}, \citenamefont {Winterer}, \citenamefont
  {Schmitz}, \citenamefont {Wende}, \citenamefont {Shvartsman},\ and\
  \citenamefont {Lupascu}}]{30}%
  \BibitemOpen
  \bibfield  {author} {\bibinfo {author} {\bibfnamefont {M.}~\bibnamefont
  {Etier}}, \bibinfo {author} {\bibfnamefont {C.}~\bibnamefont
  {Schmitz-Antoniak}}, \bibinfo {author} {\bibfnamefont {S.}~\bibnamefont
  {Salamon}}, \bibinfo {author} {\bibfnamefont {H.}~\bibnamefont {Trivedi}},
  \bibinfo {author} {\bibfnamefont {Y.}~\bibnamefont {Gao}}, \bibinfo {author}
  {\bibfnamefont {A.}~\bibnamefont {Nazrabi}}, \bibinfo {author} {\bibfnamefont
  {J.}~\bibnamefont {Landers}}, \bibinfo {author} {\bibfnamefont
  {D.}~\bibnamefont {Gautam}}, \bibinfo {author} {\bibfnamefont
  {M.}~\bibnamefont {Winterer}}, \bibinfo {author} {\bibfnamefont
  {D.}~\bibnamefont {Schmitz}}, \bibinfo {author} {\bibfnamefont
  {H.}~\bibnamefont {Wende}}, \bibinfo {author} {\bibfnamefont {V.~V.}\
  \bibnamefont {Shvartsman}}, \ and\ \bibinfo {author} {\bibfnamefont {D.~C.}\
  \bibnamefont {Lupascu}},\ }\href
  {http://www.sciencedirect.com/science/article/pii/S1359645415001329}
  {\bibfield  {journal} {\bibinfo  {journal} {Acta Materialia}\ }\textbf
  {\bibinfo {volume} {90}},\ \bibinfo {pages} {1 } (\bibinfo {year}
  {2015})}\BibitemShut {NoStop}%
\bibitem [{\citenamefont {Chu}\ \emph {et~al.}(2015)\citenamefont {Chu},
  \citenamefont {Hsu}, \citenamefont {Lu}, \citenamefont {Yang}, \citenamefont
  {yang}, \citenamefont {Luo}, \citenamefont {Yang}, \citenamefont {Hsu},
  \citenamefont {Hoffmann}, \citenamefont {Kaun},\ and\ \citenamefont
  {Lin}}]{41}%
  \BibitemOpen
  \bibfield  {author} {\bibinfo {author} {\bibfnamefont {Y.-H.}\ \bibnamefont
  {Chu}}, \bibinfo {author} {\bibfnamefont {C.-H.}\ \bibnamefont {Hsu}},
  \bibinfo {author} {\bibfnamefont {C.-I.}\ \bibnamefont {Lu}}, \bibinfo
  {author} {\bibfnamefont {H.-H.}\ \bibnamefont {Yang}}, \bibinfo {author}
  {\bibfnamefont {T.-H.}\ \bibnamefont {yang}}, \bibinfo {author}
  {\bibfnamefont {C.-H.}\ \bibnamefont {Luo}}, \bibinfo {author} {\bibfnamefont
  {K.-J.}\ \bibnamefont {Yang}}, \bibinfo {author} {\bibfnamefont {S.-H.}\
  \bibnamefont {Hsu}}, \bibinfo {author} {\bibfnamefont {G.}~\bibnamefont
  {Hoffmann}}, \bibinfo {author} {\bibfnamefont {C.-C.}\ \bibnamefont {Kaun}},
  \ and\ \bibinfo {author} {\bibfnamefont {M.-T.}\ \bibnamefont {Lin}},\
  }\href@noop {} {\bibfield  {journal} {\bibinfo  {journal} {ACS Nano}\
  }\textbf {\bibinfo {volume} {9}},\ \bibinfo {pages} {7} (\bibinfo {year}
  {2015})}\BibitemShut {NoStop}%
\bibitem [{\citenamefont {Szunyogh}\ \emph {et~al.}(2009)\citenamefont
  {Szunyogh}, \citenamefont {Lazarovits}, \citenamefont {Udvardi},
  \citenamefont {Jackson},\ and\ \citenamefont {Nowak}}]{29}%
  \BibitemOpen
  \bibfield  {author} {\bibinfo {author} {\bibfnamefont {L.}~\bibnamefont
  {Szunyogh}}, \bibinfo {author} {\bibfnamefont {B.}~\bibnamefont
  {Lazarovits}}, \bibinfo {author} {\bibfnamefont {L.}~\bibnamefont {Udvardi}},
  \bibinfo {author} {\bibfnamefont {J.}~\bibnamefont {Jackson}}, \ and\
  \bibinfo {author} {\bibfnamefont {U.}~\bibnamefont {Nowak}},\ }\href
  {\doibase 10.1103/PhysRevB.79.020403} {\bibfield  {journal} {\bibinfo
  {journal} {Phys. Rev. B}\ }\textbf {\bibinfo {volume} {79}},\ \bibinfo
  {pages} {020403} (\bibinfo {year} {2009})}\BibitemShut {NoStop}%
\bibitem [{\citenamefont {Gruber}\ \emph {et~al.}(2015)\citenamefont {Gruber},
  \citenamefont {Ibrahim}, \citenamefont {Boukari}, \citenamefont {Isshiki},
  \citenamefont {Joly}, \citenamefont {Peter}, \citenamefont {Studniarek},
  \citenamefont {da~Costa}, \citenamefont {Jabbar}, \citenamefont {Davesne},
  \citenamefont {Halisdemir}, \citenamefont {Chen}, \citenamefont {Arabski},
  \citenamefont {Otero}, \citenamefont {Choueikani}, \citenamefont {Chen},
  \citenamefont {Ohresser}, \citenamefont {Wulfhekel}, \citenamefont
  {Scheurer}, \citenamefont {Weber}, \citenamefont {Alouani}, \citenamefont
  {Beaurepaire},\ and\ \citenamefont {Bowen}}]{21}%
  \BibitemOpen
  \bibfield  {author} {\bibinfo {author} {\bibfnamefont {M.}~\bibnamefont
  {Gruber}}, \bibinfo {author} {\bibfnamefont {F.}~\bibnamefont {Ibrahim}},
  \bibinfo {author} {\bibfnamefont {S.}~\bibnamefont {Boukari}}, \bibinfo
  {author} {\bibfnamefont {H.}~\bibnamefont {Isshiki}}, \bibinfo {author}
  {\bibfnamefont {L.}~\bibnamefont {Joly}}, \bibinfo {author} {\bibfnamefont
  {M.}~\bibnamefont {Peter}}, \bibinfo {author} {\bibfnamefont
  {M.}~\bibnamefont {Studniarek}}, \bibinfo {author} {\bibfnamefont
  {V.}~\bibnamefont {da~Costa}}, \bibinfo {author} {\bibfnamefont
  {H.}~\bibnamefont {Jabbar}}, \bibinfo {author} {\bibfnamefont
  {V.}~\bibnamefont {Davesne}}, \bibinfo {author} {\bibfnamefont
  {U.}~\bibnamefont {Halisdemir}}, \bibinfo {author} {\bibfnamefont
  {J.}~\bibnamefont {Chen}}, \bibinfo {author} {\bibfnamefont {J.}~\bibnamefont
  {Arabski}}, \bibinfo {author} {\bibfnamefont {E.}~\bibnamefont {Otero}},
  \bibinfo {author} {\bibfnamefont {F.}~\bibnamefont {Choueikani}}, \bibinfo
  {author} {\bibfnamefont {K.}~\bibnamefont {Chen}}, \bibinfo {author}
  {\bibfnamefont {P.}~\bibnamefont {Ohresser}}, \bibinfo {author}
  {\bibfnamefont {W.}~\bibnamefont {Wulfhekel}}, \bibinfo {author}
  {\bibfnamefont {F.}~\bibnamefont {Scheurer}}, \bibinfo {author}
  {\bibfnamefont {W.}~\bibnamefont {Weber}}, \bibinfo {author} {\bibfnamefont
  {M.}~\bibnamefont {Alouani}}, \bibinfo {author} {\bibfnamefont
  {E.}~\bibnamefont {Beaurepaire}}, \ and\ \bibinfo {author} {\bibfnamefont
  {M.}~\bibnamefont {Bowen}},\ }\href@noop {} {\bibfield  {journal} {\bibinfo
  {journal} {Nature Materials}\ }\textbf {\bibinfo {volume} {14}},\ \bibinfo
  {pages} {981} (\bibinfo {year} {2015})}\BibitemShut {NoStop}%
\bibitem [{\citenamefont {Pike}\ and\ \citenamefont {Fernandez}(1999)}]{24}%
  \BibitemOpen
  \bibfield  {author} {\bibinfo {author} {\bibfnamefont {C.}~\bibnamefont
  {Pike}}\ and\ \bibinfo {author} {\bibfnamefont {A.}~\bibnamefont
  {Fernandez}},\ }\href {\doibase 10.1063/1.370177} {\bibfield  {journal}
  {\bibinfo  {journal} {Journal of Applied Physics}\ }\textbf {\bibinfo
  {volume} {85}},\ \bibinfo {pages} {6668} (\bibinfo {year} {1999})},\ \Eprint
  {http://arxiv.org/abs/https://doi.org/10.1063/1.370177}
  {https://doi.org/10.1063/1.370177} \BibitemShut {NoStop}%
\bibitem [{\citenamefont {Tatara}\ and\ \citenamefont {Fukuyama}(1997)}]{39}%
  \BibitemOpen
  \bibfield  {author} {\bibinfo {author} {\bibfnamefont {G.}~\bibnamefont
  {Tatara}}\ and\ \bibinfo {author} {\bibfnamefont {H.}~\bibnamefont
  {Fukuyama}},\ }\href@noop {} {\bibfield  {journal} {\bibinfo  {journal}
  {Physical Review Letters}\ }\textbf {\bibinfo {volume} {78}},\ \bibinfo
  {pages} {19} (\bibinfo {year} {1997})}\BibitemShut {NoStop}%
\bibitem [{\citenamefont {Jo}\ \emph {et~al.}(2019)\citenamefont {Jo},
  \citenamefont {Byun}, \citenamefont {Oh}, \citenamefont {Park}, \citenamefont
  {Jin}, \citenamefont {Min}, \citenamefont {Lee},\ and\ \citenamefont
  {Yoo}}]{25}%
  \BibitemOpen
  \bibfield  {author} {\bibinfo {author} {\bibfnamefont {J.}~\bibnamefont
  {Jo}}, \bibinfo {author} {\bibfnamefont {J.}~\bibnamefont {Byun}}, \bibinfo
  {author} {\bibfnamefont {I.}~\bibnamefont {Oh}}, \bibinfo {author}
  {\bibfnamefont {J.}~\bibnamefont {Park}}, \bibinfo {author} {\bibfnamefont
  {M.-J.}\ \bibnamefont {Jin}}, \bibinfo {author} {\bibfnamefont {B.-C.}\
  \bibnamefont {Min}}, \bibinfo {author} {\bibfnamefont {J.}~\bibnamefont
  {Lee}}, \ and\ \bibinfo {author} {\bibfnamefont {J.-W.}\ \bibnamefont
  {Yoo}},\ }\href {\doibase 10.1021/acsnano.8b08689} {\bibfield  {journal}
  {\bibinfo  {journal} {ACS Nano}\ }\textbf {\bibinfo {volume} {13}},\ \bibinfo
  {pages} {894} (\bibinfo {year} {2019})},\ \Eprint
  {http://arxiv.org/abs/https://doi.org/10.1021/acsnano.8b08689}
  {https://doi.org/10.1021/acsnano.8b08689} \BibitemShut {NoStop}%
\bibitem [{\citenamefont {Bisotti}\ \emph {et~al.}(2018)\citenamefont
  {Bisotti}, \citenamefont {Beg}, \citenamefont {Wang}, \citenamefont {Albert},
  \citenamefont {Chernyshenko}, \citenamefont {Cortés-Ortuño}, \citenamefont
  {Pepper}, \citenamefont {Vousden}, \citenamefont {Carey}, \citenamefont
  {Fuchs}, \citenamefont {Johansen}, \citenamefont {Balaban}, \citenamefont
  {Breth}, \citenamefont {Kluyver},\ and\ \citenamefont {Fangohr}}]{26}%
  \BibitemOpen
  \bibfield  {author} {\bibinfo {author} {\bibfnamefont {M.-A.}\ \bibnamefont
  {Bisotti}}, \bibinfo {author} {\bibfnamefont {M.}~\bibnamefont {Beg}},
  \bibinfo {author} {\bibfnamefont {W.}~\bibnamefont {Wang}}, \bibinfo {author}
  {\bibfnamefont {M.}~\bibnamefont {Albert}}, \bibinfo {author} {\bibfnamefont
  {D.}~\bibnamefont {Chernyshenko}}, \bibinfo {author} {\bibfnamefont
  {D.}~\bibnamefont {Cortés-Ortuño}}, \bibinfo {author} {\bibfnamefont
  {R.~A.}\ \bibnamefont {Pepper}}, \bibinfo {author} {\bibfnamefont
  {M.}~\bibnamefont {Vousden}}, \bibinfo {author} {\bibfnamefont
  {R.}~\bibnamefont {Carey}}, \bibinfo {author} {\bibfnamefont
  {H.}~\bibnamefont {Fuchs}}, \bibinfo {author} {\bibfnamefont
  {A.}~\bibnamefont {Johansen}}, \bibinfo {author} {\bibfnamefont
  {G.}~\bibnamefont {Balaban}}, \bibinfo {author} {\bibfnamefont
  {L.}~\bibnamefont {Breth}}, \bibinfo {author} {\bibfnamefont
  {T.}~\bibnamefont {Kluyver}}, \ and\ \bibinfo {author} {\bibfnamefont
  {H.}~\bibnamefont {Fangohr}},\ }\href {\doibase 10.5281/zenodo.1216011}
  {\enquote {\bibinfo {title} {{FinMag: finite-element micromagnetic simulation
  tool}},}\ } (\bibinfo {year} {2018})\BibitemShut {NoStop}%
\bibitem [{\citenamefont {Kresse}\ and\ \citenamefont
  {Furthmuller}(1996)}]{38}%
  \BibitemOpen
  \bibfield  {author} {\bibinfo {author} {\bibfnamefont {G.}~\bibnamefont
  {Kresse}}\ and\ \bibinfo {author} {\bibfnamefont {J.}~\bibnamefont
  {Furthmuller}},\ }\href@noop {} {\bibfield  {journal} {\bibinfo  {journal}
  {Physical Review B}\ }\textbf {\bibinfo {volume} {54}},\ \bibinfo {pages}
  {11169} (\bibinfo {year} {1996})}\BibitemShut {NoStop}%
\end{thebibliography}%

\end{document}